\documentclass[showpacs,aps,prc,preprint,nofootinbib,superscriptaddress]{revtex4-1}

\usepackage{amsmath}
\usepackage{amssymb}
\usepackage{slashed}
\usepackage{changes}
\def\bm{\boldsymbol}
\newcommand{\bea}{\begin{eqnarray}}
\newcommand{\eea}{\end{eqnarray}}
\newcommand{\be}{\begin{eqnarray}}
\newcommand{\ee}{\end{eqnarray}}
\newcommand{\no}{\nonumber \\}

\newcommand{\del}{\partial}

\def\vp{{\bm p}}

\def\vs{{\bm\sigma}}
\def\la{\langle}
\def\ra{\rangle}

\newcommand{\sixjsymbol}[6]{\left\{\begin{tabular}{ccc} {$#1$}&{$#2$}&{$#3$}\\
                             {$#4$}&{$#5$}&{$#6$} \end{tabular}\right\}}

\begin{document}


\title { Time Reversal Invariance Violating and Parity Conserving effects in Neutron Deuteron  Scattering}


\author{Young-Ho Song}
\email[]{song25@mailbox.sc.edu}
\affiliation{Department of Physics and Astronomy, University of South Carolina, Columbia, SC, 29208}

\author{Rimantas Lazauskas}
\email[]{rimantas.lazauskas@ires.in2p3.fr}
\affiliation{IPHC, IN2P3-CNRS/Universit\'e Louis Pasteur BP 28,
F-67037 Strasbourg Cedex 2, France}

\author{Vladimir Gudkov}
\email[]{gudkov@sc.edu}
\affiliation{Department of Physics and Astronomy, University of South Carolina, Columbia, SC, 29208}




\date{\today}

\begin{abstract}
Time reversal invariance violating parity conserving effects   for low energy
elastic neutron deuteron scattering
are calculated for meson exchange and EFT-type of potentials in a Distorted Wave Born Approximation,
using realistic hadronic wave functions, obtained by solving
three-body Faddeev equations in configuration space.
\end{abstract}


\maketitle

\section{Introduction
\label{sec:Intro}}

We consider phenomenological time reversal invariant violating (TRIV) and parity conserving (PC) interaction, to which we  refer  as TVPC interaction in order to distinguish it from TVPV interactions, which violate both time reversal invariance and parity. TVPC interaction  was  introduced for the first time in paper \cite{Okun:1965tu} as a possible explanation of CP-violation in $K^0$-meson decay. According to CPT theorem, the violation of CP invariance implies the TRIV. In spite of the fact that almost all known possible mechanisms of CP violation also violate parity, TVPC interactions
have been a subject of experimental and theoretical studies for decades (see, for example,  \cite{Blin-Stoyle,Kabir:1985tc,Barabanov:1986sz,Gudkov:1988ms,Moskalev:1989tc,Gudkov:1990du,Khriplovich:1990ef,Gudkov:1991qc,Gudkov:1991qg,Haxton:1993dt,Haxton:1994bq,Huffman:1996ix,PhysRevD.53.5112,Davis:1998ut,RamseyMusolf:2000wq,Kurylov:2000ub,Barabanov:2005tc} and references therein), because they directly manifest phenomena
beyond the Standard model. The most experimental constraints for this interactions were obtained by using low energy nuclear physics processes, which cover a large variety of nuclear reactions and nuclear decays.
There are a number of advantages of the search for TRIV in that processes: for example,  the possibility of enhancement of T-violating observables in neutron indused reactions by many orders of a magnitude  due to complex nuclear structure  (see, i.e. paper \cite{Gudkov:1991qg} and references therein), similar to the  enhancement observed for parity violating effects.  Another advantage to be mentioned is the existence of observables which cannot be imitated  by final state interactions \cite{BG:FSI,Kabir:1988ma,Gudkov:1992vs}. Then,  the measurement of non-zero value for  these observables  directly indicate  TRIV, similar to the case of neutron electric dipole measurements.

A promising  process for a search for TRIV in nuclear reactions is a measurement of TVPC effects in transmission of polarized neutron through polarized target \cite{Kabir:1985tc,Barabanov:1986sz}. These effects  can be  enhanced \cite{Gudkov:1991qc,Barabanov:2005tc} by a factor of $10^6$, and therefore,  could be measured at new  spallation neutron facilities, such as the SNS at the Oak Ridge National Laboratory or the J-SNS at J-PARC, Japan.

However,despite the advantage of the enhancement,
complexity of nuclear system makes it difficult to directly
relate observation of TRIV effects to nucleon TVPC coupling constants.
Therefore, it is interesting to compare
the calculations of TVPC effects in complex nuclei with
the  calculations of these effects in simplest few body systems,   which could be useful for clarification of influence of nuclear structure on values of TVPC effects. Thus, as a first step to  many body nuclear effects, we study TRIV and parity violating effects in one of the simplest available nuclear process, namely elastic neutron-deuteron scattering.
  The calculations of these effects for a specific type of TVPC interaction in a short-range approximation \cite{Gudkov:1990du}  show strong dependence of TVPC observables on neutron energy, which gives the opportunity to improve existing constrains on TVPC interactions using simple few-body system. Therefore, it is desirable to calculate these effects for general case of TVPC interactions to clarify this opportunity.

In this paper we treat TVPC nucleon-nucleon  interactions as a  perturbation,
while non-perturbed  three-body  wave functions are obtained by solving
 Faddeev equations  for realistic strong interaction Hamiltonian, based on
 AV18+UIX interaction model.
For description of TRIV potentials, we use both meson exchange model
and effective field theory (EFT) approach.
For different cases of symmetry violation we use following notations: $\slashed{T}P$ for TVPC, $\slashed{T}\slashed{P}$
for TVPV,  and
$\slashed{P}$ for parity violating (PV) cases.

\section{Observables}
\label{sec:observables}
We consider TVPC effects related to $\vs_n\cdot[{\vp}\times{\bm I}](\vp\cdot{\bm I})$ correlation, where  $\vs_n$ is the neutron spin, ${\bm I}$ is the target spin
and $\vp$ is the neutron momentum, which can be observed in the transmission of polarized neutrons through a aligned (or tensor polarized)  target.
 This correlation leads to the
difference \cite{Kabir:1985tc,Barabanov:1986sz} between the total neutron cross sections  for neutron polarized perpendicular to the neutron momentum and
averaged over elongated target states
\bea
\Delta\sigma^{\slashed{T}P}=\frac{4\pi}{p}{\rm Im}(f_{+}-f_{-}),
\eea
and neutron spin rotation angle \cite{Gudkov:1988ms,Gudkov:1991qc} $\phi$  around the axis
$[{\vp}\times{\bm I}]({\vp}\cdot{\bm I})$,
\bea
\frac{d\phi^{\slashed{T} P}}{dz}=-\frac{2\pi N}{p}{\rm Re}(f_{+}-f_{-}).
\eea
Here, $f_{+,-}$ are the zero angle scattering amplitudes for neutrons polarized parallel and anti-parallel to  the $[{\vp}\times{\bm I}](\vp\cdot{\bm I})$ axis
  respectively, $z$ is the target length, and $N$ is a number of target nuclei per unit volume.
It should be noted that for a non-zero value of this five-fold correlation, the spin $I$ must be larger or equal to one, i.e.  these TVPC effects require a tensor polarized target. Therefore, this correlation cannot be observed in nucleon-nucleon scattering.

The scattering amplitudes can be represented in terms of matrix $\hat{R}$ which is related to
scattering matrix $\hat{S}$ as $\hat{R}=\hat{1}-\hat{S}$.
In partial wave basis, we define
$R^{J}_{l^\prime {\cal S}^\prime  ,l {\cal S}}=\la l^\prime {\cal S}^\prime|R^{J}|l {\cal S}\ra$,
where unprimed and  primed parameters correspond to initial and final states,
$l$ is an orbital angular momentum
between neutron and deuteron,  ${\cal S}$ is a sum of
neutron spin and deuteron total angular momentum, and $J$ is the total angular momentum of the neutron-deuteron system.
Since we are interested in low energy neutrons, one can consider only $s$, $p$, and $d$ partial waves with mixing only between  $s$ and $d$, and $p$ and $p$ waves.  Then, one can write the TVPC parameters as
\bea
\label{eq:nsrTVPC}
\frac{1}{N}\frac{d\phi^{\slashed{T}P}}{dz}
&=&\frac{1}{2}\frac{\pi}{10p^2}\mbox{Re}
 \left[5\sqrt{2} R^{\frac{1}{2}}_{0\frac{1}{2},2\frac{3}{2}}
      -5\sqrt{2} R^{\frac{1}{2}}_{2\frac{3}{2},0\frac{1}{2}}
      +5\sqrt{2} R^{\frac{1}{2}}_{1\frac{1}{2},1\frac{3}{2}}
      -5\sqrt{2} R^{\frac{1}{2}}_{1\frac{3}{2},1\frac{1}{2}}
      \right.\no & & \left.\qquad
      +10 R^{\frac{3}{2}}_{0\frac{3}{2},2\frac{1}{2}}
      -10 R^{\frac{3}{2}}_{2\frac{1}{2},0\frac{3}{2}}
      -2\sqrt{5} R^{\frac{3}{2}}_{1\frac{1}{2},1\frac{3}{2}}
      +2\sqrt{5} R^{\frac{3}{2}}_{1\frac{3}{2},1\frac{1}{2}}
 \right],
\eea
\bea
\label{eq:dsTVPC}
\Delta\sigma^{\slashed{T} P}&=&-\frac{1}{2}\frac{\pi}{5 p^2}\mbox{Im}
 \left[5\sqrt{2} R^{\frac{1}{2}}_{0\frac{1}{2},2\frac{3}{2}}
      -5\sqrt{2} R^{\frac{1}{2}}_{2\frac{3}{2},0\frac{1}{2}}
      +5\sqrt{2} R^{\frac{1}{2}}_{1\frac{1}{2},1\frac{3}{2}}
      -5\sqrt{2} R^{\frac{1}{2}}_{1\frac{3}{2},1\frac{1}{2}}
      \right.\no & & \left.\qquad
      +10 R^{\frac{3}{2}}_{0\frac{3}{2},2\frac{1}{2}}
      -10 R^{\frac{3}{2}}_{2\frac{1}{2},0\frac{3}{2}}
      -2\sqrt{5} R^{\frac{3}{2}}_{1\frac{1}{2},1\frac{3}{2}}
      +2\sqrt{5} R^{\frac{3}{2}}_{1\frac{3}{2},1\frac{1}{2}}
 \right].
\eea

The symmetry violating $\hat{R}$ -matrix elements can be calculated with a high level of accuracy
in Distorted Wave Born Approximation (DWBA) as
\begin{equation}
\label{eq:MatEl}
R^J_{l' {\cal S}',l{\cal S}}
\simeq 4 i^{-l'+l+1} \mu p \;
{}^{(-)} \la \Psi,(l'{\cal S}')J J^z|V^{\slashed{T} P}|\Psi,(l{\cal S})J J^z\ra^{(+)},
\end{equation}
where  $\mu$ is a neutron-deuteron
reduced mass, $V^{\slashed{T}P}$ is TVPC nucleon-nucleon potential, and  $|\Psi,(l'{\cal S}')J J^z\rangle^{(\pm)}$
are  solutions of  3-body Faddeev equations in configuration space
for strong interaction Hamiltonian satisfying
outgoing(incoming) boundary condition.
The factor $i^{-l'+l}$ in this expression
is introduced to match the $R$-matrix definition
in the modified spherical harmonics convention \cite{Varshalovich}
with the  wave functions expressed
in spherical harmonics convention.
The matrix elements of TVPC potential in spherical harmonics convention
and $R$-matrix in modified spherical harmonics convention
are antisymmetric under the exchange between initial
and final states.

For calculations of wave functions, we used jj-coupling scheme instead of $l {\cal S}$ coupling scheme.
We can relate $R$-matrix elements in $l {\cal S}$ coupling scheme to  jj-coupling scheme using unitary transformation (see, for example \cite{Song:2010sz})
\bea
|[l_y\otimes(s_k \otimes j_x)_{\cal S}]_{J J_z}\rangle
&=&\sum_{j_y}|[j_x\otimes (l_y \otimes s_k)_{j_y}]_{J J_z}\rangle\no
& & \times (-1)^{j_x+j_y-J}(-1)^{l_y+s_k+j_x+J}
 [(2 j_y+1)(2 {\cal S}+1)]^{\frac{1}{2}}
 \sixjsymbol{l_y}{s_k}{j_y}{j_x}{J}{{\cal S}},
\eea
where $j_x$ is a spin (total angular momentum of the target, $j_x=1$ for the deutron),
$s_k$ is a spin of the projectile ($s_k=\frac{1}{2}$ for the neutron). In $l{\cal S}$ coupling scheme
spins of the projectile and the target are added giving partial spin ${\cal S}$ to which
relative projectile-target angular momentum ($l$) is added to obtain total angular momentum ($J$) of the system.
In contrary, in $jj$-coupling scheme, the  relative projectile-target angular momentum ($l$) is added
to the projectile spin ($s_k$) giving intermediate angular momentum ($j$) before coupling it with
target spin ($j_x$) in order to obtain total angular momentum ($J$) of the system.

\section{Time reversal violating Parity Conserving potentials
\label{sec:TVPCpot}}

The most general form of time reversal violating
and parity conserving part of nucleon-nucleon Hamiltonian
in the first order of relative nucleon momentum
can be written as \cite{Pherzeg66},
\bea
\label{eq:TVPC:general}
H^{\slashed{T}P}
&=&\left( g_1(r)+g_2(r)\tau_1\cdot\tau_2+g_3(r)T_{12}^z+g_4(r)\tau_{+}\right)
                 \hat{r}\cdot\frac{\bar\vp}{m_N}
   \no & &
   +\left(g_5(r)+g_6(r)\tau_1\cdot\tau_2+g_7(r)T_{12}^z+g_8(r)\tau_{+}\right)
                          \vs_1\cdot\vs_2\hat{r}\cdot\frac{\bar\vp}{m_N}
   \no & &
   +\left(g_9(r)+g_{10}(r)\tau_1\cdot\tau_2
        +g_{11}(r)T_{12}^z+ g_{12}(r)\tau_{+}\right)
    \no & &\quad \times
    \left(\hat{r}\cdot\vs_1\frac{\bar\vp}{m_N}\cdot\vs_2
         +\hat{r}\cdot\vs_2\frac{\bar\vp}{m_N}\cdot\vs_1
         -\frac{2}{3}\hat{r}\cdot\frac{\bar\vp}{m_N}\vs_1\cdot\vs_2
    \right)
    \no & &
   +(g_{13}(r)+g_{14}(r)\tau_1\cdot\tau_2
     +g_{15}(r)T_{12}^z+g_{16}(r)\tau_{+})
     \no & &\quad \times
    \left( \hat{r}\cdot\vs_1\hat{r}\cdot\vs_2
           \hat{r}\cdot\frac{\bar\vp}{m_N}
    -\frac{1}{5}(\hat{r}\cdot\frac{\bar\vp}{m_N}\vs_1\cdot\vs_2
                +\hat{r}\cdot\vs_1\frac{\bar{\vp}}{m_N}\cdot\vs_2
                +\hat{r}\cdot\vs_2\frac{\bar{\vp}}{m_N}\cdot\vs_1)
    \right)
    \no & &
   +g_{17}(r)
    \tau_{-}\hat{r}\cdot (\vs_\times\times\frac{\bar\vp}{m_N})
   +g_{18}(r)\tau_\times^z
    \hat{r}\cdot (\vs_{-}\times \frac{\bar\vp}{m_N}),
\eea
where exact form of
$g_i(r)$ depends on the details of a particular theory of TVPC.

One should note, that pions, being spin zero
 particles, do not contribute to $TVPC$ on-shell interaction~\cite{Simonius:1975ve}.
Therefore to describe TVPC nucleon-nucleon interactions in meson exchange potential model,
by assuming CPT conservation, one should consider contribution from heavier mesons:
$\rho(770), I^G(J^{PC})=1^+(1^{--})$ and $h_1(1170), I^G(J^{PC})=0^-(1^{+-})$ (see, for example~\cite{Haxton:1993dt,Haxton:1994bq,Gudkov:1991qc} and references therein).
For example,  Lagrangians for $\rho$ and $h_1$ are
\bea
{\cal L}^{st}&=&-g_{\rho}
  {\bar N}(\gamma_\mu \rho^{\mu,a}-\frac{\kappa_V}{2M}\sigma_{\mu\nu}
  \del^\nu \rho^{\mu,a})\tau^a N
  -g_{h}{\bar N}\gamma^\mu\gamma_5 h_\mu N,
\eea
\bea
{\cal L}^{\slashed{T}P}&=&-\frac{\bar{g}_\rho}{2m_N}
        {\bar N}\sigma^{\mu\nu}\epsilon^{3ab}\tau^a \del_\nu \rho^b_{\mu} N
        +i\frac{\bar{g}_h}{2m_N}
       {\bar N}\sigma^{\mu\nu}\gamma_5\del_\nu
       h_\mu N,
\eea
where we neglected terms, such as
${\bar N}\gamma_5\del^\mu h_\mu N$, which are small at low energy,
and $g$ and $\bar{g}$ represent strong and TVPC meson nucleon couplings
respectively.
 Then, one can obtain TVPC potentials
\bea
V^{\slashed{T}P}_\rho&=&\frac{g_\rho{\bar g}_\rho m_\rho^2}{8\pi m_N} Y_1(m_\rho r)
                \tau_{\times}^z \hat{r}\cdot
                (\vs_{-}\times\frac{\bar\vp}{m_N}),\no
V^{\slashed{T}P}_{h_1}&=&-\frac{g_h\bar{g}_h m_h^2}{2\pi m_N}Y_1(m_h r)
                  (\vs_1\cdot\frac{\bar\vp}{m_N}\vs_2\cdot\hat{r}
                  +\vs_2\cdot\frac{\bar\vp}{m_N}\vs_1\cdot\hat{r}),
\eea
where $Y_1(x)=(1+\frac{1}{x})\frac{e^-x}{x}$,  $x_a=m_a r$.
Comparing  these potentials with eq. (\ref{eq:TVPC:general}), one can see that in this model, all
$g_i(r)^{ME}=0$, except for
\bea
g_{5}^{ME}(r)&=& \left(-\frac{4 g_{h}\bar{g}_h}{3m_N}\right)
                  \left(\frac{m_h^2}{4\pi} Y_1(m_h r)\right)
              = C_{5,h}^{\slashed{T}P} f_{5,h}^{\slashed{T}P}(r,\mu=m_h),\no
g_{9}^{ME}(r)&=&\left(-\frac{2 g_{h}\bar{g}_h}{m_N}\right)
                  \left(\frac{m_h^2}{4\pi} Y_1(m_h r)\right)
              =C_{9,h}^{\slashed{T}P} f_{9,h}^{\slashed{T}P}(r,\mu=m_h)    ,\no
g_{18}^{ME}(r)&=&\left(\frac{g_{\rho}\bar{g}_\rho}{2 m_N}\right)
                  \left(\frac{m_\rho^2}{4\pi} Y_1(m_\rho r)\right)
              =C_{18,\rho}^{\slashed{T}P} f_{18,\rho}^{\slashed{T}P}(r,\mu=m_\rho) ,
\eea
where we introduced dimensional constants $C_n^{\slashed{T}P}$
and scalar function $f_n^{\slashed{T}P}(\mu)=\frac{\mu^2}{4\pi}Y_1(\mu r)$,
so that $g_n(r)$ can be written as
\bea
g_n(r)=\sum_{a} C_{n,a}^{\slashed{T}P} f_{n,a}^{\slashed{T}P}(r).
\eea
If we include  iso-vector $J=1$ $a_1$ and $b_1$ mesons,  which  masses are close to the value of $h_1$ mass, functions  $g_{6}$ and $g_{10}$ will also contribute to TVPC potential.

In EFT approach we consider eq.(\ref{eq:TVPC:general}) as a leading order  of TVPC potential.
Final result must not depend on the particular form of the $g_n(r)$ functions as long as they
are localized, like delta function or its derivative.
 Therefore,  using EFT one can estimate the contribution of each term of the potential substituting $g_n(r)$ functions by the corresponding products of low energy constants (LECs) and Yukawa functions $Y_1(\mu r)$. The mass scale $\mu$ represents a regularization scale of EFT.
For example, at low energy we can assume that $\mu\simeq m_\pi$ for pionless EFT approach.

\section{Results and discussions
\label{sec:results:tvpc}}

For calculation of TRIV amplitudes in DWBA approach, we used
the non-perturbed (time reversal invariance conserving) 3-body wave functions for neutron-deuteron scattering
 obtained by solving Faddeev equations (also often called Kowalski-Noyes
equations) in configuration space~\cite{Faddeev:1960su,Lazauskas:2004hq}.
The detailed procedure for these calculations is described in  our papers \cite{Song:2010sz,Song:2011sw}.
As previously we employed $AV18$ nucleon-nucleon potential in conjunction with $UIX$ three-nucleon force.
Obtained contribution of each TVPC operator from eq. (\ref{eq:TVPC:general})
to the matrix element of eq.(\ref{eq:MatEl}) is summarized in Tables \ref{tbl:tvpc:rho:re} and \ref{tbl:tvpc:rho:im},
representing real and imaginary parts respectively.
The  matrix elements are evaluated using $jj$ coupling scheme for the neutron-deutron center of mass energy equal $E_{cm}=100\; keV$
and the regularization scale set by $\mu=m_\rho$, which is equal to the mass of the lightest meson contributing to TVPC interaction.
It should be noted that each matrix element presented in these tables contains  a sum of contributions from different Faddeev components of wave functions with a
large number of partial waves. Therefore, the values of the matrix elements are strongly dependent on  detailed behavior of exact wave functions.  However, in spite of the fact of the possibility of a  numerical suppression of matrix elements  for some operators,  the calculated values   are stable enough to be used for estimations of TRIV effects.

\begin{table}
\caption{\label{tbl:tvpc:rho:re} Representative contribution of each
TVPC potential term to the real part of the matrix element
($\frac{1}{C_n}{\rm  Re}\frac{\la  (l' j'),J|V^{\slashed{T}P}_n|(l j),J\ra}{p^2}$).
Results are presented using $jj$-coupling scheme for wave functions obtained
using $AV18+UIX$ interaction at $E_{cm}=100$ keV.
For all operators a scalar function  $\frac{m_\rho^2}{4\pi}Y_1(m_\rho r)$ has
been used.
All data are in $fm^3$.} 
\begin{ruledtabular}
\begin{tabular}{lrrrrr}
n & $\la 2\frac{3}{2}|v^{\frac{1}{2}}|0\frac{1}{2}\ra/p^2$ & $\la 1\frac{3}{2}|v^{\frac{1}{2}}|1\frac{1}{2}\ra/p^2$
    & $\la 2\frac{3}{2}|v^{\frac{3}{2}}|0\frac{1}{2}\ra/p^2$ & $\la 1\frac{3}{2}|v^{\frac{3}{2}}|1\frac{1}{2}\ra/p^2$
   & $\la 2\frac{5}{2}|v^{\frac{3}{2}}|0\frac{1}{2}\ra/p^2$ \\
\hline
 $ 1$ & $ 0.642\times 10^{-6}$ & $ 0.184\times 10^{-6}$ & $ 0.232\times 10^{-5}$ & $ 0.342\times 10^{-6}$ & $-0.176\times 10^{-5}$ \\
 $ 2$ & $ 0.601\times 10^{-4}$ & $ 0.567\times 10^{-4}$ & $ 0.510\times 10^{-5}$ & $-0.304\times 10^{-4}$ & $ 0.346\times 10^{-4}$ \\
 $ 3$ & $-0.418\times 10^{-6}$ & $-0.614\times 10^{-6}$ & $-0.120\times 10^{-6}$ & $ 0.323\times 10^{-6}$ & $-0.335\times 10^{-6}$ \\
 $ 4$ & $-0.203\times 10^{-4}$ & $-0.185\times 10^{-4}$ & $-0.392\times 10^{-5}$ & $ 0.954\times 10^{-5}$ & $-0.952\times 10^{-5}$ \\
 $ 5$ & $-0.655\times 10^{-4}$ & $-0.582\times 10^{-4}$ & $-0.608\times 10^{-5}$ & $ 0.305\times 10^{-4}$ & $-0.283\times 10^{-4}$ \\
 $ 6$ & $-0.689\times 10^{-5}$ & $ 0.124\times 10^{-5}$ & $-0.414\times 10^{-5}$ & $-0.573\times 10^{-5}$ & $ 0.686\times 10^{-5}$ \\
 $ 7$ & $ 0.122\times 10^{-5}$ & $ 0.187\times 10^{-5}$ & $ 0.330\times 10^{-6}$ & $-0.103\times 10^{-5}$ & $ 0.100\times 10^{-5}$ \\
 $ 8$ & $ 0.668\times 10^{-4}$ & $ 0.563\times 10^{-4}$ & $ 0.719\times 10^{-5}$ & $-0.278\times 10^{-4}$ & $ 0.252\times 10^{-4}$ \\
 $ 9$ & $ 0.388\times 10^{-3}$ & $-0.262\times 10^{-3}$ & $-0.136\times 10^{-2}$ & $ 0.119\times 10^{-3}$ & $-0.714\times 10^{-3}$ \\
 $10$ & $-0.114\times 10^{-2}$ & $ 0.789\times 10^{-3}$ & $ 0.411\times 10^{-2}$ & $-0.359\times 10^{-3}$ & $ 0.214\times 10^{-2}$ \\
 $11$ & $ 0.139\times 10^{-6}$ & $ 0.837\times 10^{-8}$ & $ 0.265\times 10^{-8}$ & $ 0.166\times 10^{-7}$ & $-0.268\times 10^{-7}$ \\
 $12$ & $-0.532\times 10^{-5}$ & $-0.899\times 10^{-6}$ & $-0.638\times 10^{-5}$ & $-0.245\times 10^{-8}$ & $ 0.446\times 10^{-6}$ \\
 $13$ & $-0.307\times 10^{-4}$ & $ 0.104\times 10^{-4}$ & $ 0.835\times 10^{-4}$ & $-0.412\times 10^{-5}$ & $ 0.407\times 10^{-4}$ \\
 $14$ & $ 0.935\times 10^{-4}$ & $-0.312\times 10^{-4}$ & $-0.251\times 10^{-3}$ & $ 0.128\times 10^{-4}$ & $-0.122\times 10^{-3}$ \\
 $15$ & $ 0.170\times 10^{-7}$ & $ 0.565\times 10^{-9}$ & $ 0.338\times 10^{-9}$ & $-0.934\times 10^{-9}$ & $-0.321\times 10^{-9}$ \\
 $16$ & $-0.435\times 10^{-6}$ & $-0.630\times 10^{-7}$ & $ 0.156\times 10^{-6}$ & $-0.176\times 10^{-6}$ & $ 0.162\times 10^{-6}$ \\
 $17$ & $ 0.118\times 10^{-5}$ & $-0.274\times 10^{-4}$ & $-0.221\times 10^{-5}$ & $-0.496\times 10^{-4}$ & $ 0.536\times 10^{-5}$ \\
 $18$ & $ 0.346\times 10^{-5}$ & $-0.242\times 10^{-4}$ & $-0.257\times 10^{-5}$ & $-0.442\times 10^{-4}$ & $ 0.701\times 10^{-6}$ \\
\end{tabular}
\end{ruledtabular}
\end{table}

\begin{table}
\caption{\label{tbl:tvpc:rho:im}  Same as in Table~\ref{tbl:tvpc:rho:re}
but for the imaginary part of the matrix element
($\frac{1}{C_n}{\rm  Im}\frac{\la  (l' j'),J|V^{\slashed{T}P}_n|(l j),J\ra}{p^2}$).}
\begin{ruledtabular}
\begin{tabular}{lrrrrr}
n & $\la 2\frac{3}{2}|v^{\frac{1}{2}}|0\frac{1}{2}\ra/p^2$ & $\la 1\frac{3}{2}|v^{\frac{1}{2}}|1\frac{1}{2}\ra/p^2$
    & $\la 2\frac{3}{2}|v^{\frac{3}{2}}|0\frac{1}{2}\ra/p^2$ & $\la 1\frac{3}{2}|v^{\frac{3}{2}}|1\frac{1}{2}\ra/p^2$
   & $\la 2\frac{5}{2}|v^{\frac{3}{2}}|0\frac{1}{2}\ra/p^2$ \\
\hline
  $ 1$ & $ 0.126\times 10^{-4}$ & $-0.142\times 10^{-4}$ & $ 0.633\times 10^{-5}$ & $-0.210\times 10^{-4}$ & $-0.480\times 10^{-5}$ \\
  $ 2$ & $ 0.117\times 10^{-2}$ & $-0.437\times 10^{-2}$ & $ 0.139\times 10^{-4}$ & $ 0.187\times 10^{-2}$ & $ 0.947\times 10^{-4}$ \\
  $ 3$ & $-0.820\times 10^{-5}$ & $ 0.473\times 10^{-4}$ & $-0.329\times 10^{-6}$ & $-0.199\times 10^{-4}$ & $-0.916\times 10^{-6}$ \\
  $ 4$ & $-0.399\times 10^{-3}$ & $ 0.143\times 10^{-2}$ & $-0.107\times 10^{-4}$ & $-0.587\times 10^{-3}$ & $-0.260\times 10^{-4}$ \\
  $ 5$ & $-0.128\times 10^{-2}$ & $ 0.449\times 10^{-2}$ & $-0.166\times 10^{-4}$ & $-0.187\times 10^{-2}$ & $-0.773\times 10^{-4}$ \\
  $ 6$ & $-0.135\times 10^{-3}$ & $-0.955\times 10^{-4}$ & $-0.112\times 10^{-4}$ & $ 0.353\times 10^{-3}$ & $ 0.187\times 10^{-4}$ \\
  $ 7$ & $ 0.241\times 10^{-4}$ & $-0.144\times 10^{-3}$ & $ 0.903\times 10^{-6}$ & $ 0.638\times 10^{-4}$ & $ 0.275\times 10^{-5}$ \\
  $ 8$ & $ 0.131\times 10^{-2}$ & $-0.434\times 10^{-2}$ & $ 0.196\times 10^{-4}$ & $ 0.170\times 10^{-2}$ & $ 0.688\times 10^{-4}$ \\
  $ 9$ & $ 0.761\times 10^{-2}$ & $ 0.202\times 10^{-1}$ & $-0.372\times 10^{-2}$ & $-0.737\times 10^{-2}$ & $-0.195\times 10^{-2}$ \\
  $10$ & $-0.225\times 10^{-1}$ & $-0.608\times 10^{-1}$ & $ 0.112\times 10^{-1}$ & $ 0.221\times 10^{-1}$ & $ 0.586\times 10^{-2}$ \\
  $11$ & $ 0.273\times 10^{-5}$ & $-0.645\times 10^{-6}$ & $ 0.719\times 10^{-8}$ & $-0.102\times 10^{-5}$ & $-0.732\times 10^{-7}$ \\
  $12$ & $-0.104\times 10^{-3}$ & $ 0.693\times 10^{-4}$ & $-0.174\times 10^{-4}$ & $ 0.147\times 10^{-6}$ & $ 0.120\times 10^{-5}$ \\
  $13$ & $-0.603\times 10^{-3}$ & $-0.808\times 10^{-3}$ & $ 0.228\times 10^{-3}$ & $ 0.253\times 10^{-3}$ & $ 0.111\times 10^{-3}$ \\
  $14$ & $ 0.183\times 10^{-2}$ & $ 0.241\times 10^{-2}$ & $-0.685\times 10^{-3}$ & $-0.792\times 10^{-3}$ & $-0.335\times 10^{-3}$ \\
  $15$ & $ 0.334\times 10^{-6}$ & $-0.436\times 10^{-7}$ & $ 0.922\times 10^{-9}$ & $ 0.574\times 10^{-7}$ & $-0.878\times 10^{-9}$ \\
  $16$ & $-0.854\times 10^{-5}$ & $ 0.485\times 10^{-5}$ & $ 0.426\times 10^{-6}$ & $ 0.108\times 10^{-4}$ & $ 0.445\times 10^{-6}$ \\
  $17$ & $ 0.231\times 10^{-4}$ & $ 0.211\times 10^{-2}$ & $-0.602\times 10^{-5}$ & $ 0.305\times 10^{-2}$ & $ 0.146\times 10^{-4}$ \\
  $18$ & $ 0.678\times 10^{-4}$ & $ 0.187\times 10^{-2}$ & $-0.701\times 10^{-5}$ & $ 0.272\times 10^{-2}$ & $ 0.191\times 10^{-5}$ \\
\end{tabular}
\end{ruledtabular}
\end{table}

\begin{table}
\caption{\label{tbl:tvpc:scatt:100}
 Difference of scattering amplitudes,
$\frac{1}{C_n}\frac{(f_{n,+}-f_{n,-})}{p}$ for TVPC potential
from each operators and mass scales at $E_{cm}=100$ keV.
Note that pion mass scale does not corresponds to physical meson exchange potential.
All data are in $fm$.
}
\begin{ruledtabular}
\begin{tabular}{lrrr}
  n  &  $\frac{\Delta f^{\pi}}{p}$  &  $ \frac{\Delta f^{\rho}}{p}$ &  $\frac{\Delta f^{h_1}}{p}$ \\
\hline
  $ 1$ & $ 0.16\times 10^{-4}-i0.85\times 10^{-6}$ & $ 0.62\times 10^{-6}-i0.55\times 10^{-7}$ & $-0.25\times 10^{-6}-i0.18\times 10^{-7 }$ \\
  $ 2$ & $-0.45\times 10^{-2}-i0.39\times 10^{-4}$ & $-0.18\times 10^{-3}-i0.36\times 10^{-5}$ & $-0.53\times 10^{-4}-i0.10\times 10^{-5 }$ \\
  $ 3$ & $ 0.37\times 10^{-4}+i0.26\times 10^{-6}$ & $ 0.21\times 10^{-5}+i0.29\times 10^{-7}$ & $ 0.74\times 10^{-6}+i0.10\times 10^{-7 }$ \\
  $ 4$ & $ 0.15\times 10^{-2}+i0.14\times 10^{-4}$ & $ 0.58\times 10^{-4}+i0.12\times 10^{-5}$ & $ 0.17\times 10^{-4}+i0.35\times 10^{-6 }$ \\
  $ 5$ & $ 0.49\times 10^{-2}+i0.16\times 10^{-3}$ & $ 0.18\times 10^{-3}+i0.41\times 10^{-5}$ & $ 0.53\times 10^{-4}+i0.11\times 10^{-5 }$ \\
  $ 6$ & $-0.24\times 10^{-2}+i0.23\times 10^{-4}$ & $-0.19\times 10^{-4}+i0.32\times 10^{-6}$ & $-0.21\times 10^{-5}+i0.68\times 10^{-7 }$ \\
  $ 7$ & $-0.14\times 10^{-3}-i0.14\times 10^{-5}$ & $-0.66\times 10^{-5}-i0.91\times 10^{-7}$ & $-0.23\times 10^{-5}-i0.31\times 10^{-7 }$ \\
  $ 8$ & $-0.40\times 10^{-2}-i0.16\times 10^{-3}$ & $-0.17\times 10^{-3}-i0.42\times 10^{-5}$ & $-0.51\times 10^{-4}-i0.11\times 10^{-5 }$ \\
  $ 9$ & $ 0.42\times 10^{-1}-i0.20\times 10^{-2}$ & $ 0.14\times 10^{-2}-i0.73\times 10^{-4}$ & $ 0.44\times 10^{-3}-i0.22\times 10^{-4 }$ \\
  $10$ & $-0.12\times 10^{+0}+i0.63\times 10^{-2}$ & $-0.43\times 10^{-2}+i0.22\times 10^{-3}$ & $-0.13\times 10^{-2}+i0.68\times 10^{-4 }$ \\
  $11$ & $ 0.61\times 10^{-5}-i0.91\times 10^{-6}$ & $ 0.93\times 10^{-7}-i0.73\times 10^{-8}$ & $ 0.62\times 10^{-8}-i0.11\times 10^{-8 }$ \\
  $12$ & $-0.58\times 10^{-3}-i0.80\times 10^{-4}$ & $-0.96\times 10^{-6}+i0.11\times 10^{-6}$ & $-0.15\times 10^{-7}+i0.22\times 10^{-7 }$ \\
  $13$ & $ 0.95\times 10^{-3}-i0.41\times 10^{-4}$ & $-0.71\times 10^{-4}+i0.49\times 10^{-5}$ & $-0.25\times 10^{-4}+i0.17\times 10^{-5 }$ \\
  $14$ & $-0.23\times 10^{-2}+i0.93\times 10^{-4}$ & $ 0.22\times 10^{-3}-i0.15\times 10^{-4}$ & $ 0.75\times 10^{-4}-i0.50\times 10^{-5 }$ \\
  $15$ & $-0.52\times 10^{-6}-i0.96\times 10^{-7}$ & $ 0.16\times 10^{-7}-i0.61\times 10^{-9}$ & $ 0.43\times 10^{-8}-i0.38\times 10^{-10}$ \\
  $16$ & $-0.18\times 10^{-3}+i0.10\times 10^{-4}$ & $-0.46\times 10^{-6}+i0.28\times 10^{-7}$ & $-0.28\times 10^{-7}+i0.24\times 10^{-8 }$ \\
  $17$ & $ 0.17\times 10^{-3}+i0.16\times 10^{-4}$ & $ 0.76\times 10^{-5}+i0.64\times 10^{-8}$ & $ 0.22\times 10^{-5}-i0.38\times 10^{-8 }$ \\
  $18$ & $ 0.27\times 10^{-3}-i0.22\times 10^{-4}$ & $ 0.87\times 10^{-5}-i0.32\times 10^{-6}$ & $ 0.22\times 10^{-5}-i0.76\times 10^{-7 }$ \\
\end{tabular}
\end{ruledtabular}
\end{table}

The contributions from each TVPC operator to the difference of  scattering amplitudes  $f_{+,-}$
are summarized in Table \ref{tbl:tvpc:scatt:100}, where we distinguish three columns
representing result with a different choice of the characteristic mass scale for $g_i(r)$ functions.
 Thus, for example, the column $\frac{\Delta f^\pi}{p}$ corresponds the description of TVPC potential in pionless EFT.
(As it was mentioned above, $\pi$-meson exchange cannot lead to TVPC interaction.)

It should be noted that in spite of the fact that all results of calculations are presented only for neutron energy $E_{cm}=100\; keV$, they can be easily extrapolated for any value of neutron energy below $1\; MeV$ since  they have a simple dependence on neutron energy$E$ as:
\bea
{\rm Re}\frac{\Delta f^{\slashed{T}P}}{p}\sim \sqrt{E}, \quad
{\rm Im}\frac{\Delta f^{\slashed{T}P}}{p}\sim E.
\eea
This is because these TVPC observables are the result of a mixing of initial and final $p$-waves, or $s$- and $d$-waves, by TVPC interactions in scattering amplitudes.

To have insights into the structure of  TVPC scattering amplitudes,
one can compare them with strong,  PV, and
TVPV amplitudes at the same energy $E_{cm}=100\; keV$ ($p=0.567\times 10^{-1} \mbox{ fm}^{-1}$), and  $\mu=m_\pi$, which corresponds to pionless EFT potential.
Then,  strong $s$-wave scattering amplitude $f^{st}$ is
\bea
\frac{f^{st}}{p}=\frac{1}{2}{\rm Im}\frac{f_{+}+f_{-}}{p}
                =(-60.6+i 25.1)\; fm^2,
\eea
giving the
total cross section
$\sigma_{tot}=\frac{4\pi}{p}{\rm Im}f^{st}(p)=3.15$ b.
PV difference of scattering amplitudes in EFT is \cite{Song:2010sz}
\bea
\frac{1}{m_N C_n^{\slashed{P}}}\frac{\Delta f^{\slashed{P}}(\mu=m_\pi)}{p}
=[(-1.93\cdots 2.42)+i(-0.22\cdots 0.67)]\; fm^2,
\eea
 The difference of TRIV
amplitudes with parity violation in EFT is \cite{Song:2011sw}
\bea
\frac{1}{m_N C_n^{\slashed{T}\slashed{P}}}\frac{\Delta f^{\slashed{T}\slashed{P}}(\mu=m_\pi)}{p}
=[(-1.63\cdots 0.66)+i(-0.063\cdots 0.22)]\; fm^2,
\eea
and for TVPC ones is
\bea
\frac{1}{m_N C_n^{\slashed{T}P}}\frac{\Delta f^{\slashed{T}P}(\mu=m_\pi)}{p}
=[(-0.03\cdots 0.01)  +i(-0.0004\cdots 0.0013)]\; fm^2.
\eea
Here, $C_n^{\slashed{P}}, C_n^{\slashed{T}\slashed{P}}$, and $C_n^{\slashed{T}P}$ are  low energy constants for PV, TVPV, and TVPC interactions, correspondingly, with a separated factor $1/m_N$  to match
dimensions of the final result and retain dimension of LECs $C_n$ in $[fm]$.
The range of values for real and imaginary parts of PV and TRIV amplitudes is defined by a value of possible contribution from each PV, TVPV or TVPC operator.

Aforementioned amplitudes follow the simple kinematic rule of the suppression $\sim(p R_{nuc})$ for an additional $p$-wave involved in PV and TVPV amplitudes, and $\sim(p R_{nuc})^2$ for two $p$-waves or one $d$-wave for the case of TVPC amplitudes.
(Here, $R_{nuc}$ is an effective range of strong interaction, which leads to $\sim(p R_{nuc})\sim 0.1$ for neutron energy $E_{cm}=100\; keV$.)
In addition to this kinematic factor,  TVPC scattering amplitudes  are suppressed, as compared to PV or TVPV ones, by a factor $\frac{\bar{p}}{m_N}\sim 0.1$, which results from an extra momentum dependence of all operators in TVPC potential.  By increasing neutron energy, one can easily increase the kinematic factor up to one. Then, the only suppression of TVPC matrix elements in the amplitude will be left due to $\frac{\bar{p}}{m_N}\sim 0.1$ . It should be noted that this suppression factor is well known \cite{Gudkov:1991qc,Khriplovich:1990ef,Gudkov:1991qg,Barabanov:2005tc} for TVPC matrix elements in nuclei.

It is noteworthy that our calculations are in good agreement with results  \cite{Gudkov:1990du}, obtained using zero range force approximation for calculations of TVPC effects in $n-d$ scattering. For example, using eq.(8) of paper \cite{Gudkov:1990du} one can obtain for $E_{cm}=100\; keV$
\bea
\frac{\Delta f^{\slashed{T}P}}{p}
={g\prime}(0.0004  +i0.0013)\; fm^2,
\eea
where ${g\prime}$ is unknown TVPC nucleon-nucleon coupling constant.

The results of table \ref{tbl:tvpc:scatt:100} can also be used to express TVPC parameters in terms of
 meson exchange model. Since TVPC meson exchange model does not allow
 pion exchanges, the lightest mesons to be considered are $\rho$ and $h_1$ mesons.  Then, assuming only contributions from these mesons, one can obtain for  $E_{cm}=100\; keV$
\bea
\Delta\sigma^{\slashed{T}P}&=& 10^{-6}
 [g_h\bar{g}_h(1.15)-g_\rho \bar{g}_\rho (4.22\cdot 10^{-3})] \mbox{ b},\no
\frac{1}{N}\frac{d\phi^{\slashed{T}P}}{dz}&=&10^{-3}
 [g_h\bar{g}_h(1.25 )-g_\rho\bar{g}_\rho(5.76\cdot 10^{-3})] \mbox{ rad fm}^2 .
\eea

Finally, we  conclude that neutron-deuteron scattering is a promising process to improve current experimental constraints on TVPC interactions.
The TVPC
observables can be large enough to be measured at neutron energy of hundreds of $keV$ due to strong energy dependence.
On the other hand they can be precisely calculated,
providing the possibility to extract the TVPC nucleon coupling constants from the experiment.

\begin{acknowledgments}
This work was supported by the DOE grants no. DE-FG02-09ER41621.
This work was granted access to the HPC resources of IDRIS
under the allocation 2009-i2009056006
made by GENCI (Grand Equipement National de Calcul Intensif).
We thank the staff members of the IDRIS for their constant help.
\end{acknowledgments}

\bibliography{TViolation}

\begin{thebibliography}{28}%
\makeatletter
\providecommand \@ifxundefined [1]{%
 \@ifx{#1\undefined}
}%
\providecommand \@ifnum [1]{%
 \ifnum #1\expandafter \@firstoftwo
 \else \expandafter \@secondoftwo
 \fi
}%
\providecommand \@ifx [1]{%
 \ifx #1\expandafter \@firstoftwo
 \else \expandafter \@secondoftwo
 \fi
}%
\providecommand \natexlab [1]{#1}%
\providecommand \enquote  [1]{``#1''}%
\providecommand \bibnamefont  [1]{#1}%
\providecommand \bibfnamefont [1]{#1}%
\providecommand \citenamefont [1]{#1}%
\providecommand \href@noop [0]{\@secondoftwo}%
\providecommand \href [0]{\begingroup \@sanitize@url \@href}%
\providecommand \@href[1]{\@@startlink{#1}\@@href}%
\providecommand \@@href[1]{\endgroup#1\@@endlink}%
\providecommand \@sanitize@url [0]{\catcode `\\12\catcode `\$12\catcode
  `\&12\catcode `\#12\catcode `\^12\catcode `\_12\catcode `\%12\relax}%
\providecommand \@@startlink[1]{}%
\providecommand \@@endlink[0]{}%
\providecommand \url  [0]{\begingroup\@sanitize@url \@url }%
\providecommand \@url [1]{\endgroup\@href {#1}{\urlprefix }}%
\providecommand \urlprefix  [0]{URL }%
\providecommand \Eprint [0]{\href }%
\providecommand \doibase [0]{http://dx.doi.org/}%
\providecommand \selectlanguage [0]{\@gobble}%
\providecommand \bibinfo  [0]{\@secondoftwo}%
\providecommand \bibfield  [0]{\@secondoftwo}%
\providecommand \translation [1]{[#1]}%
\providecommand \BibitemOpen [0]{}%
\providecommand \bibitemStop [0]{}%
\providecommand \bibitemNoStop [0]{.\EOS\space}%
\providecommand \EOS [0]{\spacefactor3000\relax}%
\providecommand \BibitemShut  [1]{\csname bibitem#1\endcsname}%
\let\auto@bib@innerbib\@empty
\bibitem [{\citenamefont {Okun}(1965)}]{Okun:1965tu}%
  \BibitemOpen
  \bibfield  {author} {\bibinfo {author} {\bibfnamefont {L.~B.}\ \bibnamefont
  {Okun}},\ }\href@noop {} {\bibfield  {journal} {\bibinfo  {journal} {Sov. J.
  Nucl. Phys.}\ }\textbf {\bibinfo {volume} {1}},\ \bibinfo {pages} {670}
  (\bibinfo {year} {1965})}\BibitemShut {NoStop}%
\bibitem [{\citenamefont {Blin-Stoyle}(1973)}]{Blin-Stoyle}%
  \BibitemOpen
  \bibfield  {author} {\bibinfo {author} {\bibfnamefont {R.~J.}\ \bibnamefont
  {Blin-Stoyle}},\ }\href@noop {} {\emph {\bibinfo {title} {Fundamental
  Interactions and the Nucleus}}}\ (\bibinfo  {publisher} {American Elsevier},\
  \bibinfo {year} {1973})\BibitemShut {NoStop}%
\bibitem [{\citenamefont {Kabir}()}]{Kabir:1985tc}%
  \BibitemOpen
  \bibfield  {author} {\bibinfo {author} {\bibfnamefont {P.~K.}\ \bibnamefont
  {Kabir}},\ }\href@noop {} {\ }\bibinfo {note} {In The investigation of
  fundamental interaction with cold neutrons, edited by G. L. Greene (NBS
  Special Pub., 711, 1985), p.81.}\BibitemShut {Stop}%
\bibitem [{\citenamefont {Barabanov}(1986)}]{Barabanov:1986sz}%
  \BibitemOpen
  \bibfield  {author} {\bibinfo {author} {\bibfnamefont {A.~L.}\ \bibnamefont
  {Barabanov}},\ }\href@noop {} {\bibfield  {journal} {\bibinfo  {journal}
  {Sov. J. Nucl. Phys.}\ }\textbf {\bibinfo {volume} {44}},\ \bibinfo {pages}
  {775} (\bibinfo {year} {1986})}\BibitemShut {NoStop}%
\bibitem [{\citenamefont {Gudkov}()}]{Gudkov:1988ms}%
  \BibitemOpen
  \bibfield  {author} {\bibinfo {author} {\bibfnamefont {V.~P.}\ \bibnamefont
  {Gudkov}},\ }\href@noop {} {\ }\bibinfo {note} {In Fundamental symmetries and
  nuclear structure, edited by J. N. Ginocchio and S. P. Rosen (World
  Scientific, Singapore, 1988), p.21-35.}\BibitemShut {Stop}%
\bibitem [{\citenamefont {Moskalev}\ and\ \citenamefont
  {Prsev}(1989)}]{Moskalev:1989tc}%
  \BibitemOpen
  \bibfield  {author} {\bibinfo {author} {\bibfnamefont {A.~N.}\ \bibnamefont
  {Moskalev}}\ and\ \bibinfo {author} {\bibfnamefont {S.~G.}\ \bibnamefont
  {Prsev}},\ }\href@noop {} {\bibfield  {journal} {\bibinfo  {journal} {Sov. J.
  Nucl. Phys.}\ }\textbf {\bibinfo {volume} {49}},\ \bibinfo {pages} {789}
  (\bibinfo {year} {1989})}\BibitemShut {NoStop}%
\bibitem [{\citenamefont {Gudkov}(1990)}]{Gudkov:1990du}%
  \BibitemOpen
  \bibfield  {author} {\bibinfo {author} {\bibfnamefont {V.~P.}\ \bibnamefont
  {Gudkov}},\ }\href@noop {} {\bibfield  {journal} {\bibinfo  {journal}
  {Z.Phys.}\ }\textbf {\bibinfo {volume} {A337}},\ \bibinfo {pages} {247}
  (\bibinfo {year} {1990})}\BibitemShut {NoStop}%
\bibitem [{\citenamefont {Khriplovich}(1991)}]{Khriplovich:1990ef}%
  \BibitemOpen
  \bibfield  {author} {\bibinfo {author} {\bibfnamefont {I.~B.}\ \bibnamefont
  {Khriplovich}},\ }\href@noop {} {\bibfield  {journal} {\bibinfo  {journal}
  {Nucl. Phys.}\ }\textbf {\bibinfo {volume} {B352}},\ \bibinfo {pages} {385}
  (\bibinfo {year} {1991})}\BibitemShut {NoStop}%
\bibitem [{\citenamefont {Gudkov}(1991)}]{Gudkov:1991qc}%
  \BibitemOpen
  \bibfield  {author} {\bibinfo {author} {\bibfnamefont {V.~P.}\ \bibnamefont
  {Gudkov}},\ }\href@noop {} {\bibfield  {journal} {\bibinfo  {journal}
  {Nucl.Phys.}\ }\textbf {\bibinfo {volume} {A524}},\ \bibinfo {pages} {668}
  (\bibinfo {year} {1991})}\BibitemShut {NoStop}%
\bibitem [{\citenamefont {Gudkov}(1992{\natexlab{a}})}]{Gudkov:1991qg}%
  \BibitemOpen
  \bibfield  {author} {\bibinfo {author} {\bibfnamefont {V.~P.}\ \bibnamefont
  {Gudkov}},\ }\href@noop {} {\bibfield  {journal} {\bibinfo  {journal} {Phys.
  Rept.}\ }\textbf {\bibinfo {volume} {212}},\ \bibinfo {pages} {77} (\bibinfo
  {year} {1992}{\natexlab{a}})}\BibitemShut {NoStop}%
\bibitem [{\citenamefont {Haxton}\ and\ \citenamefont
  {Horing}(1993)}]{Haxton:1993dt}%
  \BibitemOpen
  \bibfield  {author} {\bibinfo {author} {\bibfnamefont {W.~C.}\ \bibnamefont
  {Haxton}}\ and\ \bibinfo {author} {\bibfnamefont {A.}~\bibnamefont
  {Horing}},\ }\href@noop {} {\bibfield  {journal} {\bibinfo  {journal} {Nucl.
  Phys.}\ }\textbf {\bibinfo {volume} {A560}},\ \bibinfo {pages} {469}
  (\bibinfo {year} {1993})}\BibitemShut {NoStop}%
\bibitem [{\citenamefont {Haxton}\ \emph {et~al.}(1994)\citenamefont {Haxton},
  \citenamefont {Horing},\ and\ \citenamefont {Musolf}}]{Haxton:1994bq}%
  \BibitemOpen
  \bibfield  {author} {\bibinfo {author} {\bibfnamefont {W.~C.}\ \bibnamefont
  {Haxton}}, \bibinfo {author} {\bibfnamefont {A.}~\bibnamefont {Horing}}, \
  and\ \bibinfo {author} {\bibfnamefont {M.~J.}\ \bibnamefont {Musolf}},\
  }\href@noop {} {\bibfield  {journal} {\bibinfo  {journal} {Phys. Rev.}\
  }\textbf {\bibinfo {volume} {D50}},\ \bibinfo {pages} {3422} (\bibinfo {year}
  {1994})}\BibitemShut {NoStop}%
\bibitem [{\citenamefont {Huffman}\ \emph {et~al.}(1997)\citenamefont {Huffman}
  \emph {et~al.}}]{Huffman:1996ix}%
  \BibitemOpen
  \bibfield  {author} {\bibinfo {author} {\bibfnamefont {P.~R.}\ \bibnamefont
  {Huffman}} \emph {et~al.},\ }\href@noop {} {\bibfield  {journal} {\bibinfo
  {journal} {Phys. Rev.}\ }\textbf {\bibinfo {volume} {C55}},\ \bibinfo {pages}
  {2684} (\bibinfo {year} {1997})}\BibitemShut {NoStop}%
\bibitem [{\citenamefont {Engel}\ \emph {et~al.}(1996)\citenamefont {Engel},
  \citenamefont {Frampton},\ and\ \citenamefont {Springer}}]{PhysRevD.53.5112}%
  \BibitemOpen
  \bibfield  {author} {\bibinfo {author} {\bibfnamefont {J.}~\bibnamefont
  {Engel}}, \bibinfo {author} {\bibfnamefont {P.~H.}\ \bibnamefont {Frampton}},
  \ and\ \bibinfo {author} {\bibfnamefont {R.~P.}\ \bibnamefont {Springer}},\
  }\href@noop {} {\bibfield  {journal} {\bibinfo  {journal} {Phys. Rev. D}\
  }\textbf {\bibinfo {volume} {53}},\ \bibinfo {pages} {5112} (\bibinfo {year}
  {1996})}\BibitemShut {NoStop}%
\bibitem [{\citenamefont {Davis}\ and\ \citenamefont
  {Gould}(1999)}]{Davis:1998ut}%
  \BibitemOpen
  \bibfield  {author} {\bibinfo {author} {\bibfnamefont {E.~D.}\ \bibnamefont
  {Davis}}\ and\ \bibinfo {author} {\bibfnamefont {C.~R.}\ \bibnamefont
  {Gould}},\ }\href@noop {} {\bibfield  {journal} {\bibinfo  {journal} {Phys.
  Lett.}\ }\textbf {\bibinfo {volume} {B447}},\ \bibinfo {pages} {209}
  (\bibinfo {year} {1999})}\BibitemShut {NoStop}%
\bibitem [{\citenamefont {Ramsey-Musolf}()}]{RamseyMusolf:2000wq}%
  \BibitemOpen
  \bibfield  {author} {\bibinfo {author} {\bibfnamefont {M.~J.}\ \bibnamefont
  {Ramsey-Musolf}},\ }\href@noop {} {\ }\bibinfo {note} {In Workshop on
  Fundamental Physics with Pulsed Neutron Beams (FPPNB 2000), Research Triangle
  Park, North Carolina, 1-3 Jun 2000}\BibitemShut {NoStop}%
\bibitem [{\citenamefont {Kurylov}\ \emph {et~al.}(2001)\citenamefont
  {Kurylov}, \citenamefont {McLaughlin},\ and\ \citenamefont
  {Ramsey-Musolf}}]{Kurylov:2000ub}%
  \BibitemOpen
  \bibfield  {author} {\bibinfo {author} {\bibfnamefont {A.}~\bibnamefont
  {Kurylov}}, \bibinfo {author} {\bibfnamefont {G.~C.}\ \bibnamefont
  {McLaughlin}}, \ and\ \bibinfo {author} {\bibfnamefont {M.~J.}\ \bibnamefont
  {Ramsey-Musolf}},\ }\href@noop {} {\bibfield  {journal} {\bibinfo  {journal}
  {Phys. Rev.}\ }\textbf {\bibinfo {volume} {D63}},\ \bibinfo {pages} {076007}
  (\bibinfo {year} {2001})}\BibitemShut {NoStop}%
\bibitem [{\citenamefont {Barabanov}\ and\ \citenamefont
  {Beda}(2005)}]{Barabanov:2005tc}%
  \BibitemOpen
  \bibfield  {author} {\bibinfo {author} {\bibfnamefont {A.~L.}\ \bibnamefont
  {Barabanov}}\ and\ \bibinfo {author} {\bibfnamefont {A.~G.}\ \bibnamefont
  {Beda}},\ }\href@noop {} {\bibfield  {journal} {\bibinfo  {journal} {J. Phys.
  G: Nucl. Part. Phys.}\ }\textbf {\bibinfo {volume} {31}},\ \bibinfo {pages}
  {161} (\bibinfo {year} {2005})}\BibitemShut {NoStop}%
\bibitem [{\citenamefont {Bunakov}\ and\ \citenamefont
  {Gudkov}(1984)}]{BG:FSI}%
  \BibitemOpen
  \bibfield  {author} {\bibinfo {author} {\bibfnamefont {V.~E.}\ \bibnamefont
  {Bunakov}}\ and\ \bibinfo {author} {\bibfnamefont {V.~P.}\ \bibnamefont
  {Gudkov}},\ }\href@noop {} {\bibfield  {journal} {\bibinfo  {journal} {J.
  Phys.(Paris) Colloq.}\ }\textbf {\bibinfo {volume} {45}},\ \bibinfo {pages}
  {C3} (\bibinfo {year} {1984})}\BibitemShut {NoStop}%
\bibitem [{\citenamefont {Kabir}(1988)}]{Kabir:1988ma}%
  \BibitemOpen
  \bibfield  {author} {\bibinfo {author} {\bibfnamefont {P.~K.}\ \bibnamefont
  {Kabir}},\ }\href {\doibase 10.1103/PhysRevD.37.1856} {\bibfield  {journal}
  {\bibinfo  {journal} {Phys. Rev.}\ }\textbf {\bibinfo {volume} {D37}},\
  \bibinfo {pages} {1856} (\bibinfo {year} {1988})}\BibitemShut {NoStop}%
\bibitem [{\citenamefont {Gudkov}(1992{\natexlab{b}})}]{Gudkov:1992vs}%
  \BibitemOpen
  \bibfield  {author} {\bibinfo {author} {\bibfnamefont {V.~P.}\ \bibnamefont
  {Gudkov}},\ }\href@noop {} {\bibfield  {journal} {\bibinfo  {journal} {Phys.
  Rev.}\ }\textbf {\bibinfo {volume} {C46}},\ \bibinfo {pages} {357} (\bibinfo
  {year} {1992}{\natexlab{b}})}\BibitemShut {NoStop}%
\bibitem [{\citenamefont {Varshalovich}\ \emph {et~al.}(1988)\citenamefont
  {Varshalovich}, \citenamefont {Moskalev},\ and\ \citenamefont
  {Khersonskii}}]{Varshalovich}%
  \BibitemOpen
  \bibfield  {author} {\bibinfo {author} {\bibfnamefont {D.~A.}\ \bibnamefont
  {Varshalovich}}, \bibinfo {author} {\bibfnamefont {A.~N.}\ \bibnamefont
  {Moskalev}}, \ and\ \bibinfo {author} {\bibfnamefont {V.~K.}\ \bibnamefont
  {Khersonskii}},\ }\href@noop {} {\emph {\bibinfo {title} {Quantum Theory of
  Angular Momentum}}}\ (\bibinfo  {publisher} {World Scientific},\ \bibinfo
  {year} {1988})\BibitemShut {NoStop}%
\bibitem [{\citenamefont {Song}\ \emph
  {et~al.}(2011{\natexlab{a}})\citenamefont {Song}, \citenamefont {Lazauskas},\
  and\ \citenamefont {Gudkov}}]{Song:2010sz}%
  \BibitemOpen
  \bibfield  {author} {\bibinfo {author} {\bibfnamefont {Y.-H.}\ \bibnamefont
  {Song}}, \bibinfo {author} {\bibfnamefont {R.}~\bibnamefont {Lazauskas}}, \
  and\ \bibinfo {author} {\bibfnamefont {V.}~\bibnamefont {Gudkov}},\
  }\href@noop {} {\bibfield  {journal} {\bibinfo  {journal} {Phys. Rev.}\
  }\textbf {\bibinfo {volume} {C83}},\ \bibinfo {pages} {015501} (\bibinfo
  {year} {2011}{\natexlab{a}})}\BibitemShut {NoStop}%
\bibitem [{\citenamefont {Herczeg}(1966)}]{Pherzeg66}%
  \BibitemOpen
  \bibfield  {author} {\bibinfo {author} {\bibfnamefont {P.}~\bibnamefont
  {Herczeg}},\ }\href@noop {} {\bibfield  {journal} {\bibinfo  {journal} {Nucl.
  Phys.}\ }\textbf {\bibinfo {volume} {75}},\ \bibinfo {pages} {655} (\bibinfo
  {year} {1966})}\BibitemShut {NoStop}%
\bibitem [{\citenamefont {Simonius}(1975)}]{Simonius:1975ve}%
  \BibitemOpen
  \bibfield  {author} {\bibinfo {author} {\bibfnamefont {M.}~\bibnamefont
  {Simonius}},\ }\href@noop {} {\bibfield  {journal} {\bibinfo  {journal}
  {Phys. Lett.}\ }\textbf {\bibinfo {volume} {B58}},\ \bibinfo {pages} {147}
  (\bibinfo {year} {1975})}\BibitemShut {NoStop}%
\bibitem [{\citenamefont {Faddeev}(1961)}]{Faddeev:1960su}%
  \BibitemOpen
  \bibfield  {author} {\bibinfo {author} {\bibfnamefont {L.~D.}\ \bibnamefont
  {Faddeev}},\ }\href@noop {} {\bibfield  {journal} {\bibinfo  {journal} {Sov.
  Phys. JETP}\ }\textbf {\bibinfo {volume} {12}},\ \bibinfo {pages} {1014}
  (\bibinfo {year} {1961})}\BibitemShut {NoStop}%
\bibitem [{\citenamefont {Lazauskas}\ and\ \citenamefont
  {Carbonell}(2004)}]{Lazauskas:2004hq}%
  \BibitemOpen
  \bibfield  {author} {\bibinfo {author} {\bibfnamefont {R.}~\bibnamefont
  {Lazauskas}}\ and\ \bibinfo {author} {\bibfnamefont {J.}~\bibnamefont
  {Carbonell}},\ }\href@noop {} {\bibfield  {journal} {\bibinfo  {journal}
  {Phys. Rev.}\ }\textbf {\bibinfo {volume} {C70}},\ \bibinfo {pages} {044002}
  (\bibinfo {year} {2004})}\BibitemShut {NoStop}%
\bibitem [{\citenamefont {Song}\ \emph
  {et~al.}(2011{\natexlab{b}})\citenamefont {Song}, \citenamefont {Lazauskas},\
  and\ \citenamefont {Gudkov}}]{Song:2011sw}%
  \BibitemOpen
  \bibfield  {author} {\bibinfo {author} {\bibfnamefont {Y.-H.}\ \bibnamefont
  {Song}}, \bibinfo {author} {\bibfnamefont {R.}~\bibnamefont {Lazauskas}}, \
  and\ \bibinfo {author} {\bibfnamefont {V.}~\bibnamefont {Gudkov}},\
  }\href@noop {} {\  (\bibinfo {year} {2011}{\natexlab{b}})},\ \Eprint
  {http://arxiv.org/abs/1104.3051} {arXiv:1104.3051 [nucl-th]} \BibitemShut
  {NoStop}%
\end{thebibliography}%

\end{document}